\documentclass[a4paper,12pt]{article}

\usepackage{float}
\usepackage{harvard}
\citationmode{abbr} \citationstyle{dcu}
\usepackage{amsmath}
\usepackage{amssymb}
\usepackage{graphicx}
\usepackage{graphicx}
\usepackage{lscape}
\usepackage{epic}

\begin{document}


\title{A dynamical programming approach for controlling the directed abelian Dhar-Ramaswamy model}
\author{Daniel O. Cajueiro$^{1, 3}$ and Roberto F. S. Andrade$^{2, 3}$}
\date{}
\maketitle

\begin{center}$^1$Department of Economics -- Universidade de Bras\'{i}lia, DF 70910-900, Brazil.\\$^2$Instituto de F\'{i}sica, Universidade Federal da Bahia, BA 40210-340,
Brazil.\\$^3$National Institute of Science and Technology for
Complex Systems,  Brazil.\end{center}

\begin{abstract} A dynamical programming approach is used to deal with the problem
of controlling the directed abelian Dhar-Ramaswamy  model on
two-dimensional square lattice. Two strategies are considered to
obtain explicit results to this task. First, the optimal solution
of the problem is characterized by the solution of the Bellman
equation obtained by numerical algorithms. Second, the solution is
used as a benchmark to value how far from the optimum other
heuristics that can be applied to larger systems are. This
approach is the first attempt on the direction of schemes for
controlling self-organized criticality that are based on
optimization principles that consider explicitly a tradeoff
between the size of the avalanches and the cost of intervention.
\end{abstract}

\section{Introduction}

Self-organized criticality (SOC) is a characteristic of systems
that are driven by a slowly acting external force and organize
themselves through energy dissipating avalanches of all sizes.
Although SOC models were first proposed to explain the origin of
the $1/f$ noise, it is recognized that it can be used to explain a
large class of systems such as earthquakes~\cite{sch91},
evolutionary bursts~\cite{baksne93}, forest fires~\cite{drosch92},
rice piles~\cite{fre96}, financial markets~\cite{bar06}, and so
on.

Recently we have raised the issue that, for some critical
organized systems, the size of the largest avalanches can be
reduced by a control intervention heuristics~\cite{cajand10}. That
was just a first effort to investigate if and how the damaging
energy dissipative bursts in SOC systems could be controlled.
Although such systems organize themselves without external
intervention, this reorganization in a lower level of energy is
very costly for society, since it depends on avalanches of all
sizes. Examples of these events of dissipation of energy in nature
and society are avalanches that arise in snow hills, bubbles that
explode in financial markets and earthquakes. Although it is not
possible to intervene in events such as earthquakes, in some sense
we can intervene in the process that generates large snow
avalanches and the explosion of stock bubbles. In the former case,
small avalanches can be triggered in order to avoid larger
ones~\cite{mccsch93}. In the later case, central banks can in some
sense enforce a monetary policy that can avoid the rising of large
bubbles~\cite{greenspan08}. Regarding this aspect, it is not the
purpose of this work to defend this kind of procedure, but we
surely think that it deserves to be studied.  Our first
investigation ~\cite{cajand10}, was based on a replica model of
the region of the system to be controlled, a control scheme was
designed to externally trigger small size avalanches in order to
avoid larger ones. Although we have shown that this principle
works for sandpiles in two-dimensional lattices~\cite{cajand10},
we have no information about how far from the optimal choice these
heuristics are. To fill this gap we resume our investigation with
a rather different approach: we develop a dynamical programming
(DP) approach to control the directed abelian Dhar-Ramaswamy
(DADR) model in a two dimensional lattice. Due to the huge number
of possible states that come to play in DP approaches, any
feasible investigation must be restricted to systems of much
smaller size than those one usually considers when performing
numerical integration of the systems. Nevertheless, we can use
this approach to characterize optimally the problem of controlling
SOC systems, as well as to explore the efficiency of other
heuristics built without any kind of optimization law and as a
basis for approximate optimization principles such as the ones
presented in~\cite{bertsi96}.

It is worth mentioning that recent literature has used
optimization principles to understand the structure and dynamics
of several complex systems.
In~\cite{rodrin92,caj05,jacrog05,mottor07,carior08}, it was shown
that complex networks may arise from optimization principles.
Further, analyzes of optimal navigation in complex
networks~\cite{caj09,caj10} have shown that a walker that
minimizes the cost of walking overlaps the random walker and the
directed walker behaviors. In~\cite{cajmal08,ast08} optimization
has been used to study the complex human dynamics of task
execution. Moreover, reinforcement learning has been used to
explore the problem of learning paths in complex
networks~\cite{cajand09}.

\section{The DP approach for controlling the DADR model}
\label{sec:dp} The DADR  model~\cite{dha89} is built on a
two-dimensional square lattice of $L\times L$ sites $(i,j)$,
$i,j=1,\cdots,L$. Each site stores a certain amount $z_{ij}$ of
mass units. At each time step, the system is driven by two update
rules: (a) Addition rule: a mass unit is added to a randomly
selected site $(k,\ell)$, so that $z_{k\ell}\rightarrow
z_{k\ell}+1$. (b) Toppling rule: if $z_{ij}>z_{c}=1$, then
$z_{ij}\rightarrow z_{ij}-2$, $z_{i+1,j-1}\rightarrow
z_{i+1,j-1}+1$ and $z_{i+1,j+1}\rightarrow z_{i+1,j+1}+1$. The
model is usually represented after performing a $5\pi/4$ rotation
of the standard square lattice, in such a way the site $(i+1,j+1)$
lies just below the site $(i,j)$, and the $\mathbf{x}$ and
$\mathbf{y}$ directions are at $5\pi/4$ and $7\pi/4$ angles with
the horizontal axis. Therefore, if deposition occurs in site
$(i,j)$, the only sites that may be affected are those located on
the lines $i+\ell, \ell\geq 1$.

Let $\Gamma$ be the finite set of stable states (configurations)
in the phase space of the DADR model, and $N_\Gamma$ the number of
elements of $\Gamma$. As in any DP study, it is necessary to
identify the different actions (or policies) that can be taken
when the system is in any of these states. So let us note such one
policy as $\pi=[u(1),\cdots,u(N_\Gamma)]$, where $u(i)\in U$
refers to the specific control action that $\pi$ undertakes when
the system is in the state $i$. $U$ represents the set of
admissible controls, i.e. those control actions that do not
violate the system dynamics. Note that the number of elements in
the set $\Pi$ of all admissible policies $\pi$ increases faster
than combinatorial when the system size increases. Indeed, this
number depends both on $N_\Gamma$ and on the number of possible
control actions for each state $i$.

To control the avalanches sizes in our approach, it is important
to consider that events occur according to an ordered sequence, as
discussed in ~\cite{cajand10}. If $t$ is a discrete variable
$t=n\Delta t$, and the DADR model is in a given ``stable'' state
$x_t$ we assume that the following events take place within the
time step $\Delta t$. The control scheme triggers (or not) one
avalanche in a specific site $(i,j)$ of the lattice. If this
occurs, the avalanche starts by emptying the site $(i,j)$, what
amounts to topple the single unit mass with $50\%$ of probability
to the site $(i+1,j-1)$ or to the site $(i+1,j+1)$. This control
induced toppling may lead to further toppling until the system
reaches a new stable state $x_{t}^{c}$ due to the control
intervention $u(x_t)$. After this induced avalanche, which is
absent if the used policy indicates to take no action, the usual
deposition process of the model takes place and the system evolves
from either $x_t$ or $x_t^c$ to a new state $x_{t+1}$. For this
process, we only care that control intervention comes before the
deposition process, and that both relaxations occur within the
same time step $\Delta t$.

However, differently from~\cite{cajand10}, we assume here that
there is only possible at most one control intervention per time
step, and the intervention decision is made according to a dynamic
programming approach.

For instance, if $L=2$, then $N_\Gamma=2^{L^2}=2^{2^2}=16$. One
possible state of $\Gamma$ is

\begin{equation}
    x=\left
    [\begin{array}{cc}
    0 & 1 \\
    1 & 0 \\
  \end{array}
  \right].\label{aa}
\end{equation}

\noindent Please note that, for the purpose of keeping a simple
diagram, we did not perform the rotation used for the
representation of the system described before. In such matrix-like
notation, the toppling process makes the grain move either one
column to the right or one line downwards. Assume  $x$ to be
$x_t$. In this state, we have three admissible controls, namely
the one that triggers no avalanche, and those that trigger an
avalanche in the sites $(1,2)$ and $(2,1)$, respectively. If there
is no intervention, the intermediate state $x_c(t)=x(t)$. If an
avalanche is triggered in the site $(1,2)$, the system goes to
$x_{t}^{c}$ which, with equal probabilities, is one of the states

\[\left[\begin{array}{cc}
    0 & 0 \\
    1 & 1 \\
  \end{array}\right], \left[\begin{array}{cc}
    0 & 0 \\
    1 & 0 \\
  \end{array}\right].\]

\noindent (If the site $(2,1)$ had been chosen, the situation
would be quite similar but, for the sake of brevity, we will not
consider this choice here.) In order to differ one state from the
other, we call the one in the left as $x_{t}^{c,L}$ and the one in
the right as $x_{t}^{c,R}$, making reference to the side followed
by the controlled avalanche. Thus, after the deposition process,
the system can have suffered a transition to one of the following
states, in the case of $x_{t}^{c,L}$,

\[\left[\begin{array}{cc}
    1 & 0 \\
    1 & 1 \\
  \end{array}\right], \left[\begin{array}{cc}
    0 & 1 \\
    1 & 1 \\
  \end{array}\right], \left[\begin{array}{cc}
    0 & 0 \\
    0 & 0 \\
  \end{array}\right], \left[\begin{array}{cc}
    0 & 0 \\
    1 & 0 \\
  \end{array}\right].\]

\noindent or to one of the possible states, if $x_{t}^{c,R}$ was
taken:

\[\left[\begin{array}{cc}
    1 & 0 \\
    1 & 0 \\
  \end{array}\right], \left[\begin{array}{cc}
    0 & 1 \\
    1 & 0 \\
  \end{array}\right], \left[\begin{array}{cc}
    0 & 0 \\
    0 & 1 \\
  \end{array}\right], \left[\begin{array}{cc}
    0 & 0 \\
    1 & 1 \\
  \end{array}\right].\]

\noindent Therefore, with equal probabilities, the state $x_{t+1}$
will be represented by one of these eight states. One also must
note that, associated with the control intervention and the
deposition process, we have respectively two classes of
avalanches:  {\it controlled avalanches}, that are triggered by
the control scheme in state $x_t$ with sizes $s_{c,x_t\,
x_{t}^{c,L}}$ or $s_{c,x_t \, x_{t}^{c,R}}$ depending on the side
that the controlled avalanche followed; and {\it uncontrolled
avalanches} with sizes $s_{u,x_{t}^{c,L} \, x_{t+1}}$ or
$s_{u,x_{t}^{c,R} \, x_{t+1}}$ that happen when the system goes
from state $x_{t}^{c,L}$ or $x_{t}^{c,R}$ to state $x_{t+1}$ due
to the deposition process.

Following the DP approach, we assume that the control scheme makes
the decision of triggering an avalanche in one site of the system
or doing nothing in a given state $x$ based on the minimization of
the cost function

\begin{equation}J(x)= \min_{\pi\in\Pi}E^{\pi}\left[\sum_{t=0}^{\infty} \gamma^t g(x_t,u(x_t))|x\right]
\label{eq:prob}
\end{equation}

\noindent where the expectation $E^\pi[\cdot|x]$ is conditional on
the policy $\pi$ and the state $x$.  The cost per stage
$g(\cdot,\cdot)$ is given by

\begin{equation}
g(x_t,u(x_t))=\sum_{x_{t+1}}p_{x_t x_{t+1}}(u)
\overline{g}(x_t,u(x_t),x_{t+1}), \label{eq:g}
\end{equation}

\noindent where $u(x_t)$ is the control intervention in state
$x_t$, $\overline{g}(x_t,u(x_t),x_{t+1})$ is the cost of using the
control $u$ at state $x_t$ and moving to state $x_{t+1}$, $p_{x_t
x_{t+1}}(u)$ is the transition probability from state $x_t$ to
state $x_{t+1}$ using the control $u$ at state $x_t$. In the
general expression (\ref{eq:g}), the function form of
$\overline{g}$ has not yet made explicit. In this work, we assume
a simple particular form for $\overline{g}$ that depends only on
two parameters and a functional dependence on the avalanche size
$s$. Therefore, we write

\begin{eqnarray}
\lefteqn{g(x_t,u(x_t))= C_I I(x_t)}\\&& +\frac{1}{2}[ C_A
h(s_{c,x_t\, x_{t}^{c,L}}) + \sum_{x_{t+1}}p_{x_{t}^{c,L} x_{t+1}}
C_A h(s_{u,x_{t}^{c,L}\, x_{t+1}})]\nonumber \\&& +\frac{1}{2}[C_A
h(s_{c,x_t\, x_{t}^{c,R}}) + \sum_{x_{t+1}}p_{x_{t}^{c,R} x_{t+1}}
C_A h(s_{u,x_{t}^{c,R}\, x_{t+1}})], \nonumber \label{eq:gb}
\end{eqnarray}

\noindent where $p_{x_{t}^{c,L} x_{t+1}}$ (or $p_{x_{t}^{c,R}
x_{t+1}}$) is the probability of transition from $x_{t}^{c,L}$ (or
$x_{t}^{c,R}$) to $x_{t+1}$. One must also note that, while
$p_{x_t x_{t+1}}(u)$ indicates explicitly that the probability of
the transition from state $x_t$ to state $x_{t+1}$ depends on the
choice of the control $u$ at state $x_t$, $p_{x_{t}^{c,L}
x_{t+1}}$ (or $p_{x_{t}^{c,R} x_{t+1}}$) does not present this
dependence. $I(x_t)$ is an indicator function that assumes the
value 1, when there is an intervention in state $x_t$, and 0
otherwise. Finally, the two parameters $C_I$ and $C_A$ represent
the fixed cost associated with one intervention and avalanche
size. For the sake of simplicity, we assume here that $h(s)=s^2$,
i.e., we penalize larger avalanches in a power law with exponent
equal to 2.

The term  $\gamma^t$ in (\ref{eq:prob}) weights differently the
influence of the present and future costs in the decision process.
Although a realistic optimization process must take into account
the intervention cost, it could be expected that avalanches at a
given time step and those in the next future should have
approximately the same weight in the decision process, what
amounts to take the discount factor $\gamma=1$. We call the
attention that this simple choice leads to a technical difficult,
namely, we can not ensure that this problem has a solution that
does not depend on the kind of the controlled Markov process. In
such situations, the method used to solve the problem may depend
on the type of the controlled Markov chain that we are dealing
with and may be difficult to find by simple
algorithms~\cite{put05}. However, due to the Banach Fixed
Theorem~\cite{put05}, this difficult can be circumvented if we
consider $\gamma\rightarrow 1_-$. Finally we should also note that
the choice $\gamma = 1$ is somewhat unrealistic as it does not
consider the cost of the money over time.

It is easy to show that the solution of problem (\ref{eq:prob}) is
given by the Bellman equation~\cite{bel57}

\begin{equation} J(x)=\min_{u\in U(x)}\left[g(x,u)+\sum_{x'}p_{x x'}(u) \gamma J(x')\right].\label{bellman}\end{equation}

In the rest of this paper, we discuss the solution of problem
(\ref{eq:prob}) using numerical solutions of the Bellman equation
(\ref{bellman}) found by means of the value iteration
algorithm~\cite{put05}.

\section{Results} \label{sec:results}

\begin{figure}[t]
\begin{tabular}{cc}
  \includegraphics[width=4cm,height=4cm]{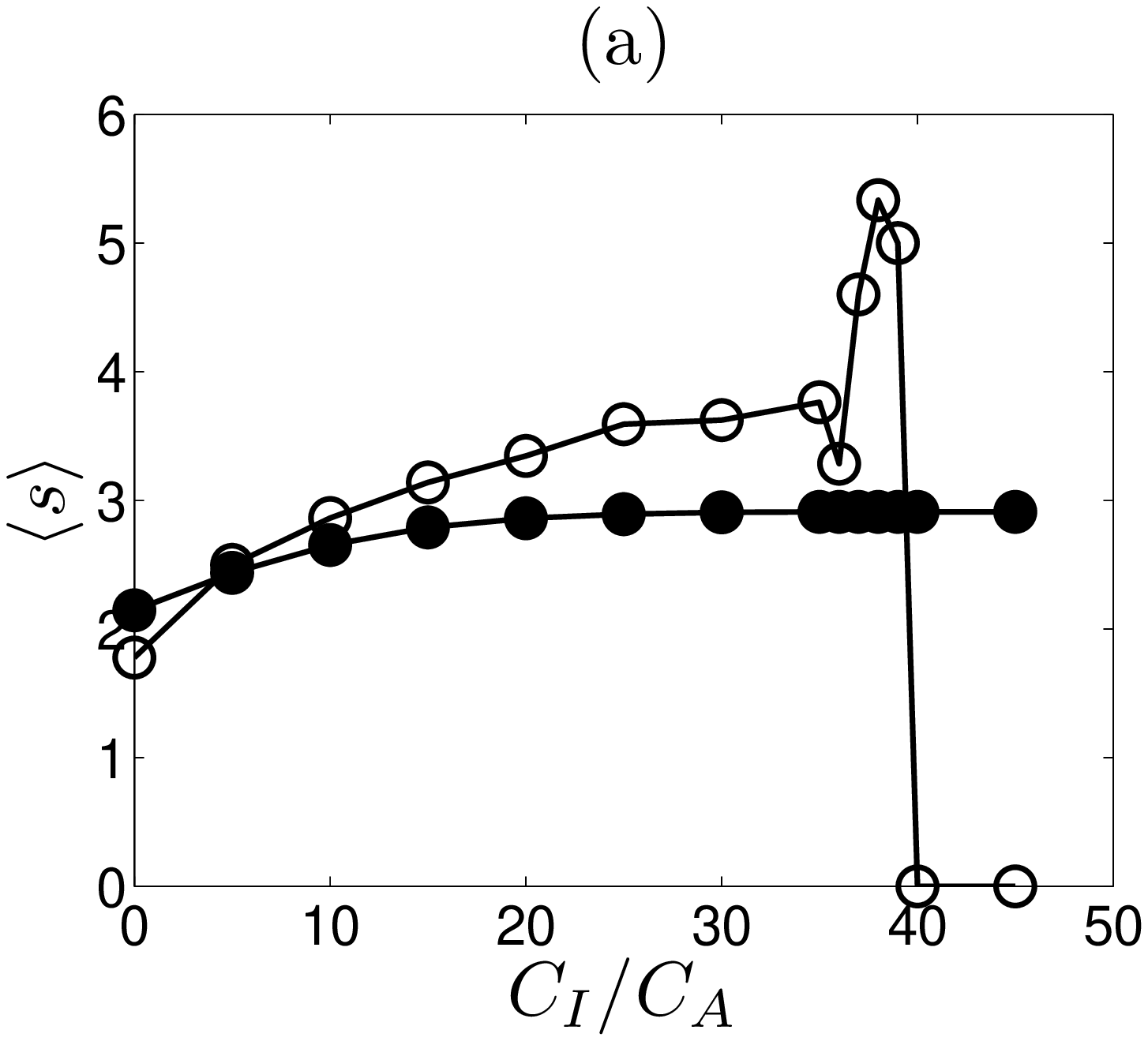} & \includegraphics[width=4cm,height=4cm]{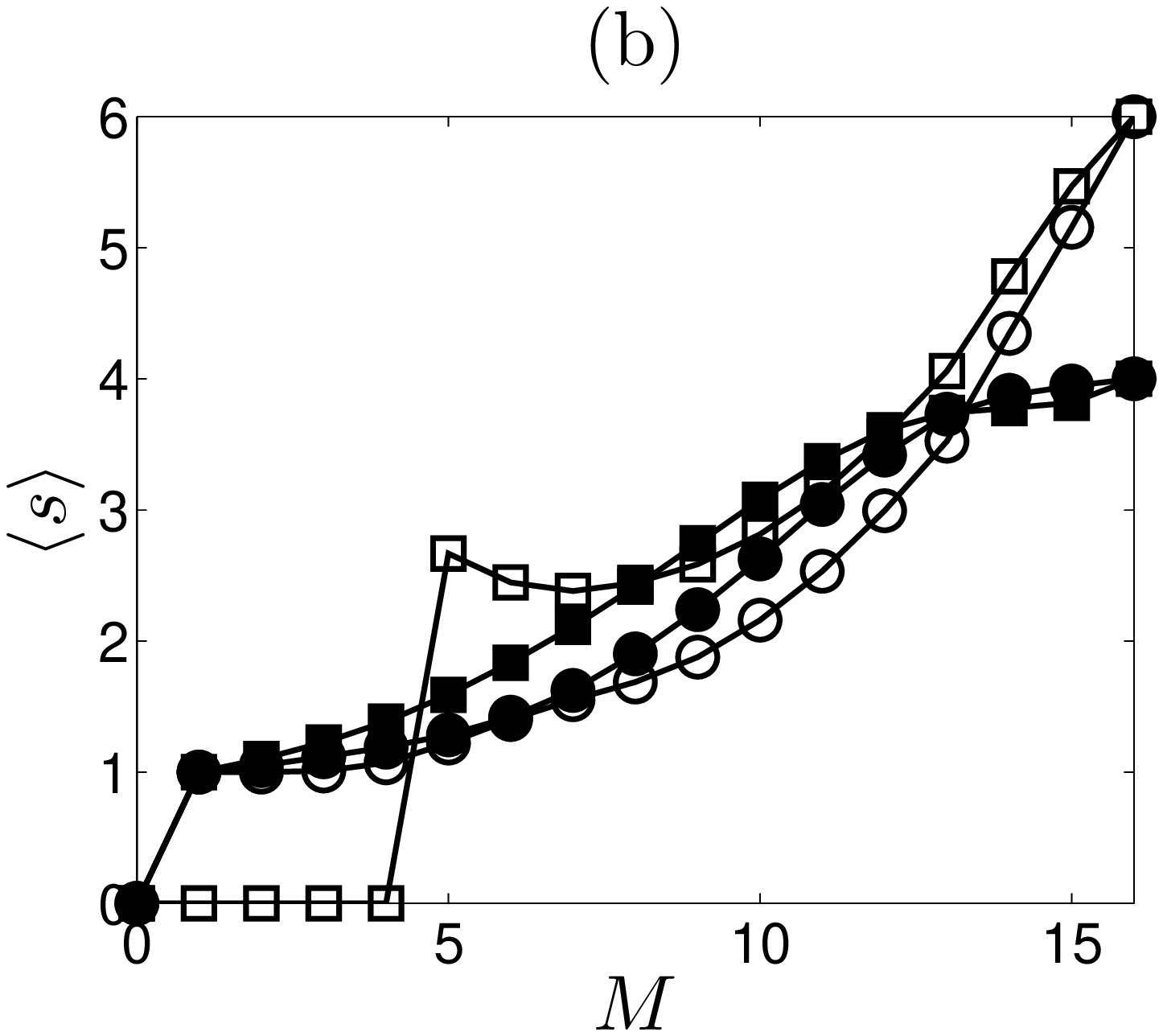} \\
\end{tabular}

\caption{(a) The avalanche average size for several values of the
ratio $C_I/C_A$: controlled avalanches (hollow circles) and
uncontrolled avalanches (solid circles). (b) The average size of
the avalanches for several values of the mass of the state. While
solid symbols represent uncontrolled avalanches, hollow symbols
represent controlled avalanches: $C_I/C_A=0$ (circles) and
$C_I/C_A=10$ (squares).} \label{figura1}
\end{figure}

The main difficulty associated with DP approach is the rapid
increase of $N_\Gamma$ which, for the current study, behaves like
$2^{L^2}$. For practical purposes, it becomes impossible to study
numerically a system larger than $L=4$. As already discussed, the
size of such system in much smaller than lattice sizes actually
used to compute the time evolution of the model. However, we will
show that this approach can be used to characterize the optimal
solution of the problem and be used as a benchmark to validate
other solutions based on ad-hoc chosen heuristics.

For all simulations of DADR's model reported in this paper, the
corresponding solution of Eq. (\ref{bellman}) was obtained for
$L=4$ and $\gamma=0.999$. For this value of $L$, $N_\Gamma=
65536$, the largest avalanche that can take place in such system
is of size $s=16$, the minimal and maximal amount of mass $M$ kept
in the system are, respectively, $M=0$ and $M=16$. For the next
lattice size $L=5$, solving (\ref{bellman}) requires to find the
minimum of $J(x)$ by taking into account all $2^{25}\sim 3.2
\times 10^7$ states for this lattice size.

At a given time $t$, the number of admissible controls depends on
the state $x$. For instance, while in the unique state of the
system with mass 0, there is only one control, which is to do
nothing, in the unique state of the system with mass 16, there are
17 admissible controls.

Figure \ref{figura1}(a) shows the effect of the the cost $C_A$ and
$C_I$ in the solution of the problem presenting the average
controlled and uncontrolled avalanches $\langle s\rangle$ for
different values of $C_I/C_A$.  It is shown that, when the cost of
making interventions becomes high, the control scheme waits until
the system has stored a larger amount of mass $M$ to make
interventions. This causes also the size of the uncontrolled
avalanches to increase. For a given threshold value
$(C_I/C_A)_T\sim 40$, intervention cost becomes so large that the
optimal solution corresponds to not intervene in the system
anymore. Correspondingly, when $C_I/C_A$ is close to
$(C_I/C_A)_T$, the number of interventions decreases exponentially
(not shown).  Figure 1a also shows that, for $C_I/C_A >
(C_I/C_A)_T$, only avalanches produced by the system dynamics are
observed.

It turns out that $M$ is an interesting metric that can be used to
order the states of the system in terms of danger of larger
avalanches. Figure \ref{figura1}(b) shows the average size of
avalanches for several values of $M$ for the ratios $C_I/C_A=0$
and $C_I/C_A=10$. This figure shows the effect of the increasing
the ratio $C_I/C_A$ for a state of the system characterized by its
mass. For the no cost intervention situation $C_I/C_A=0$, the
control scheme acts for all states of the system but the one with
$M=0$. The same does not happen for $C_I/C_A=10$, when avalanches
are triggered only for states with $M>4$. The consequence of such
behavior is to increase the size of the uncontrolled avalanches
for the states with low values of $M$. Only for large values of
$M$ the average size of the controlled avalanches becomes larger
than that of the uncontrolled ones. Moreover, one may also see
that increasing $C_I/C_A$ has almost no effect on the avalanche
average size when $M$ grows. Finally, Figure \ref{figura1}(b)
suggests that the $C_I/C_A$ plays a role similar to the acceptable
size of an avalanche considered in~\cite{cajand10}, i.e., when the
ratio $C_I/C_A$ is high, it is not worth triggering small
controlled avalanches anymore.

Now, we compare the results provided by DP control with those from
three other heuristic approaches that we identify as maximal ({\it
max}), minimal ({\it min}) and random ({\it ran}), although none
of them is exactly equivalent to the fixed avalanche size
heuristic discussed in \cite{cajand10}. As in the DP case, all of
them make at most one intervention per time step. We call $p_I$
the fraction of time steps where an intervention occurs. Let $T_I$
be a minimal threshold avalanche size that allows the {\it max}
and {\it min} control schemes to intervene, i.e., they do not
trigger avalanches with size less than $T_I$. This parameter plays
a role similar to the acceptable size of an avalanche
in~\cite{cajand10}. The {\it max} approach works as follows. In
each time step $t$ and corresponding state $x_t$, it triggers only
the maximal avalanche with size $s_{\mathrm{max}}$ that may happen
in this state if $s_{\mathrm{max}}\ge T_I$. On the other hand, the
{\it min} approach triggers the minimal avalanche with size
$s_{\mathrm{min}}$ that may happen in the state $x_t$ if
$s_{\mathrm{min}}\ge T_I$. Finally, the {\it ran} approach
triggers avalanches in saturated sites of the system with the same
frequency of intervention $p_I$.

\begin{figure}[t]
\begin{tabular}{cc}
  \includegraphics[width=4.2cm,height=4.2cm]{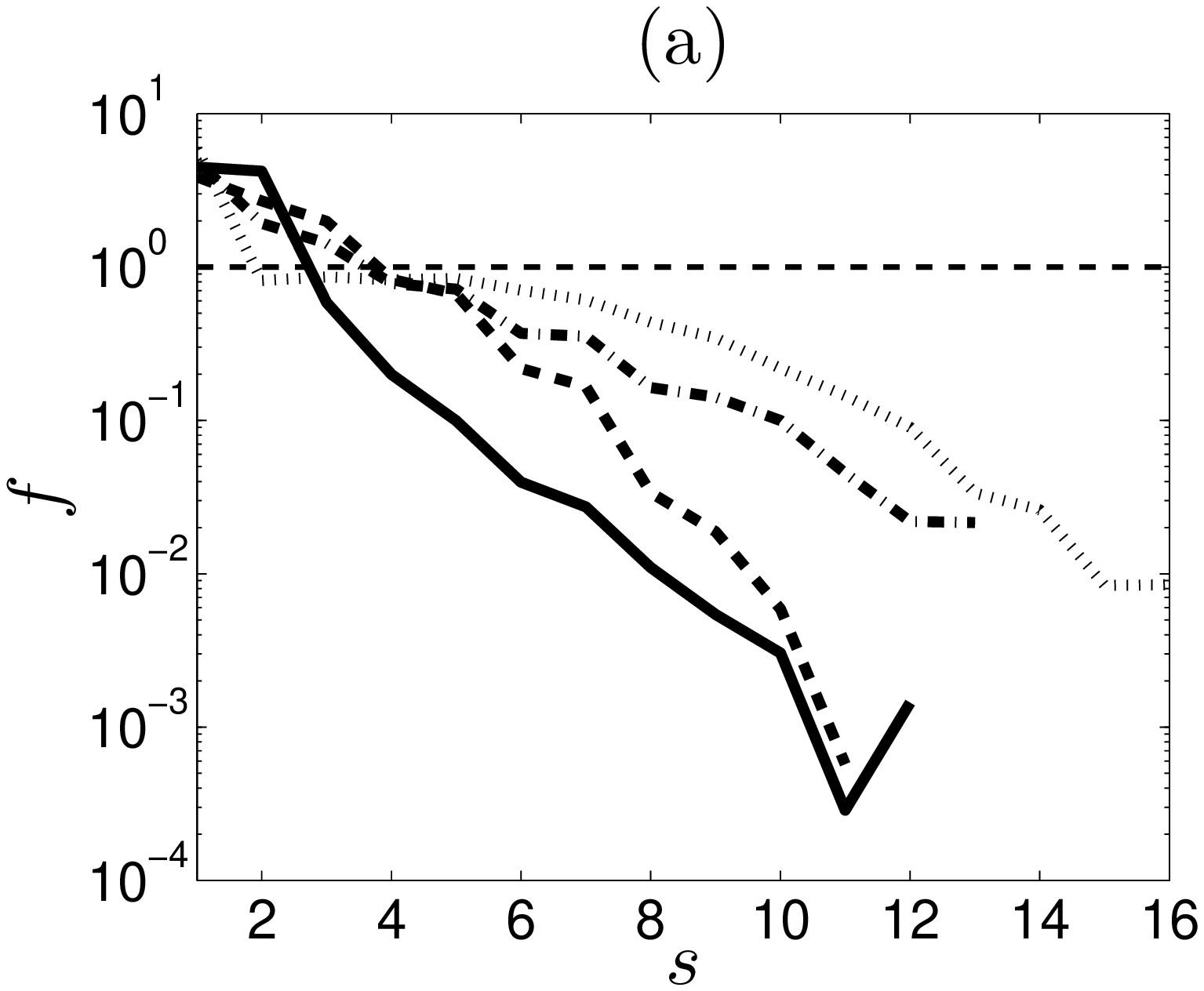} & \includegraphics[width=4.2cm,height=4.2cm]{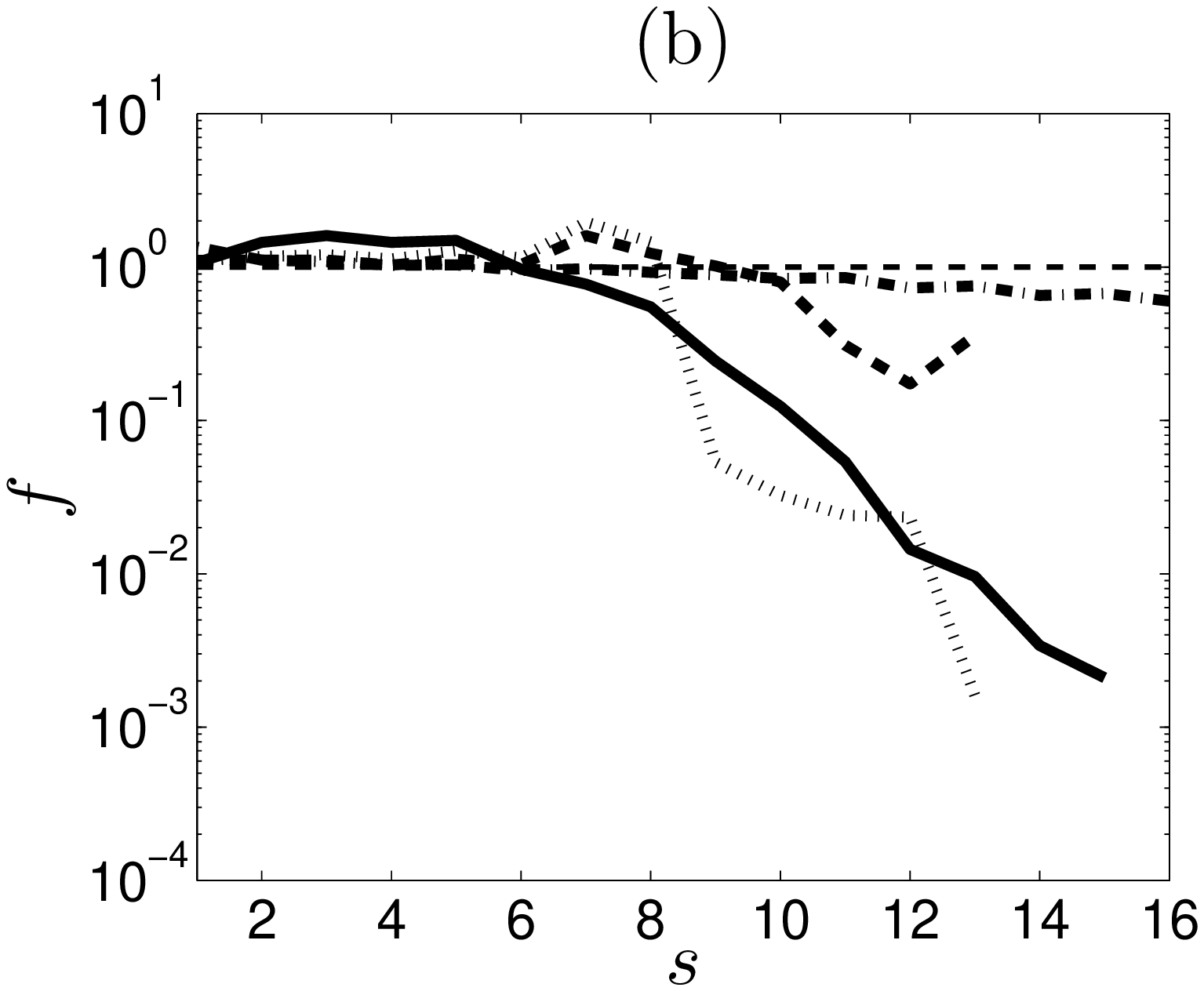} \\
\end{tabular}
\caption{Ratio $f$ between the total number of avalanches in the
controlled and uncontrolled situations, for several control
schemes. Both in panel (a) as in (b), the following notation is
used to identify parameter values and adopted control scheme:
$(p_I;\mathrm{cost\; scheme}/\mathrm{cost\; DP};
\mathrm{scheme})$. In panel (a), $T_I=1$: solid  (1; 1; DP),
dashes (1; 1.29; {\it max}), dots (1; 1.46; {\it min}), dot-dashes
(1; 1.31; {\it ran}). In panel (b), $T_I=8$: solid (0.11; 1; DP),
dashes (0.12; 1.21; {\it max}), dots (0.12; 1.14; {\it min}),
dot-dashes (0.1; 1.19; {\it ran}).}\label{figura3}
\end{figure}

In order to compare the results of the four approaches, we use the
number of interventions as a tune parameter. Therefore, we choose
$T_I$ large enough in order to have the number of interventions of
the {\it max} and {\it min} schemes of the same order of the DP
control. Figure \ref{figura3} compares the DP results low (a) and
high (b) ratios $C_I/C_A$ with the equivalent {\it max}, {\it min}
and {\it ran} controls. There we measure the efficiency of the
control scheme by the ratio $f$ between the number of avalanches
of the controlled to the uncontrolled system for $T_I=1$
(\ref{figura3}a) and $T_I=8$ (\ref{figura3}b). From these
strategies, we see that the {\it max} scheme performs more closely
to the optimal one when the control scheme is allowed to make
almost one intervention per time step and the {\it min} scheme
performs better when the control schemes are allowed only to make
interventions when there is a probability of large avalanches. It
is clear that the cost is always minimal for the DP scheme.
Furthermore, while the {\it max}, {\it min} and {\it ran} controls
have their performances strongly affected by changes in the ratio
$C_I/C_A$, the DP control makes a good work in reducing the size
of avalanches in both situations (this information can also be
seen with the help of Fig. \ref{figura1}(b)). Figure \ref{figura3}
can also help us to choose when to choose the {\it min} scheme and
the {\it max} scheme. The {\it min} scheme should be used when the
size of $T_I$ is larger -- triggering the minimal avalanches, this
system can avoid uncontrolled triggering of large avalanches. On
the other hand, for low values of $T_I$, one should use the {\it
max} scheme. Since in almost every time step avalanches are being
triggered, triggering the largest ones the {\it max} control
avoids uncontrolled triggering of larger avalanches. Note that the
use of the {\it max\;}  scheme for the case of large $T_I$ is
dangerous, since the control by itself will trigger large
avalanches. Finally, the choice of the {\it min} control for small
$T_I$ is useless, since it will trigger only small avalanches that
do not help avoid the largest ones.

\begin{figure}[t]
\begin{tabular}{cc}
  \includegraphics[width=4.2cm,height=4.2cm]{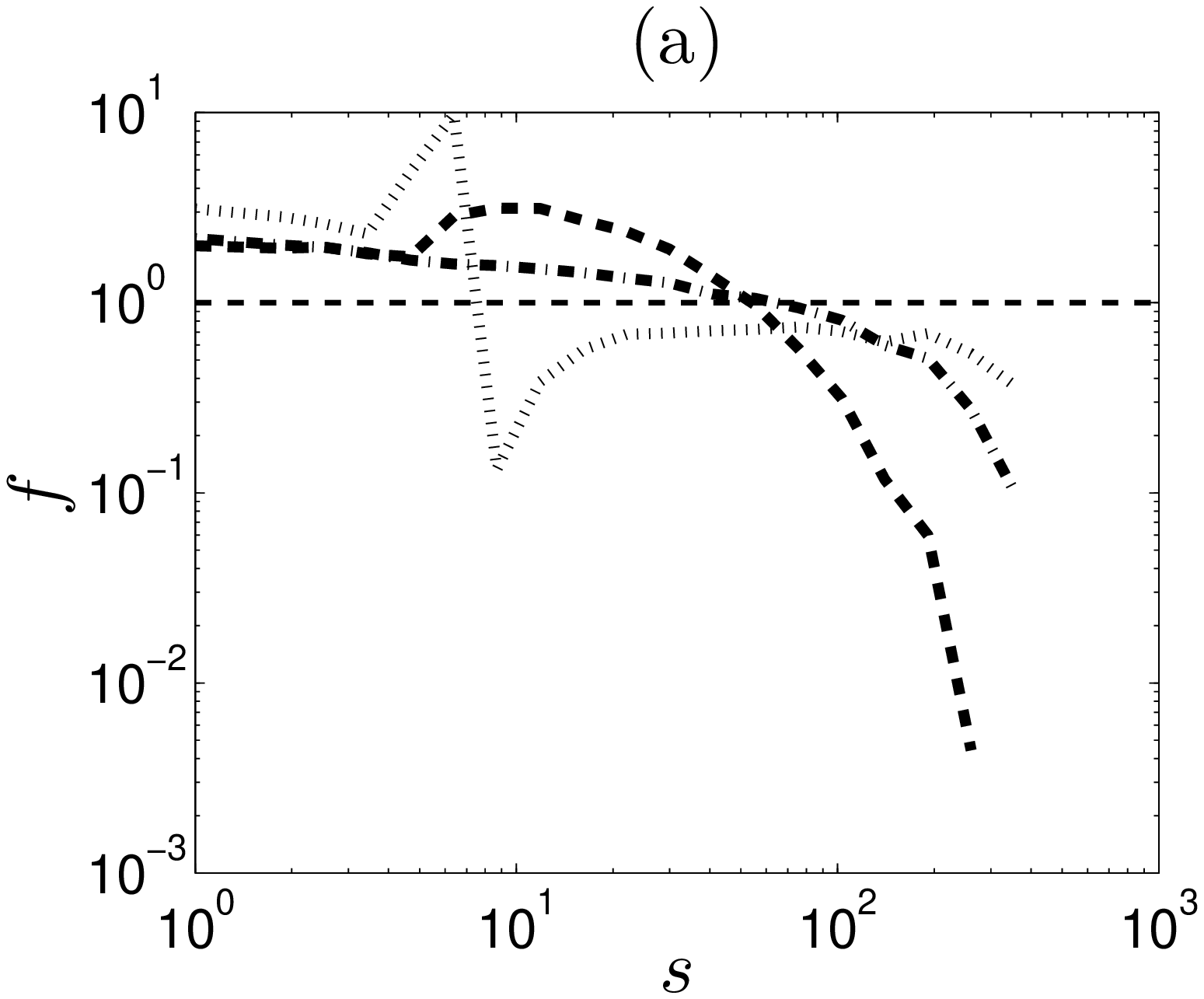} & \includegraphics[width=4.2cm,height=4.2cm]{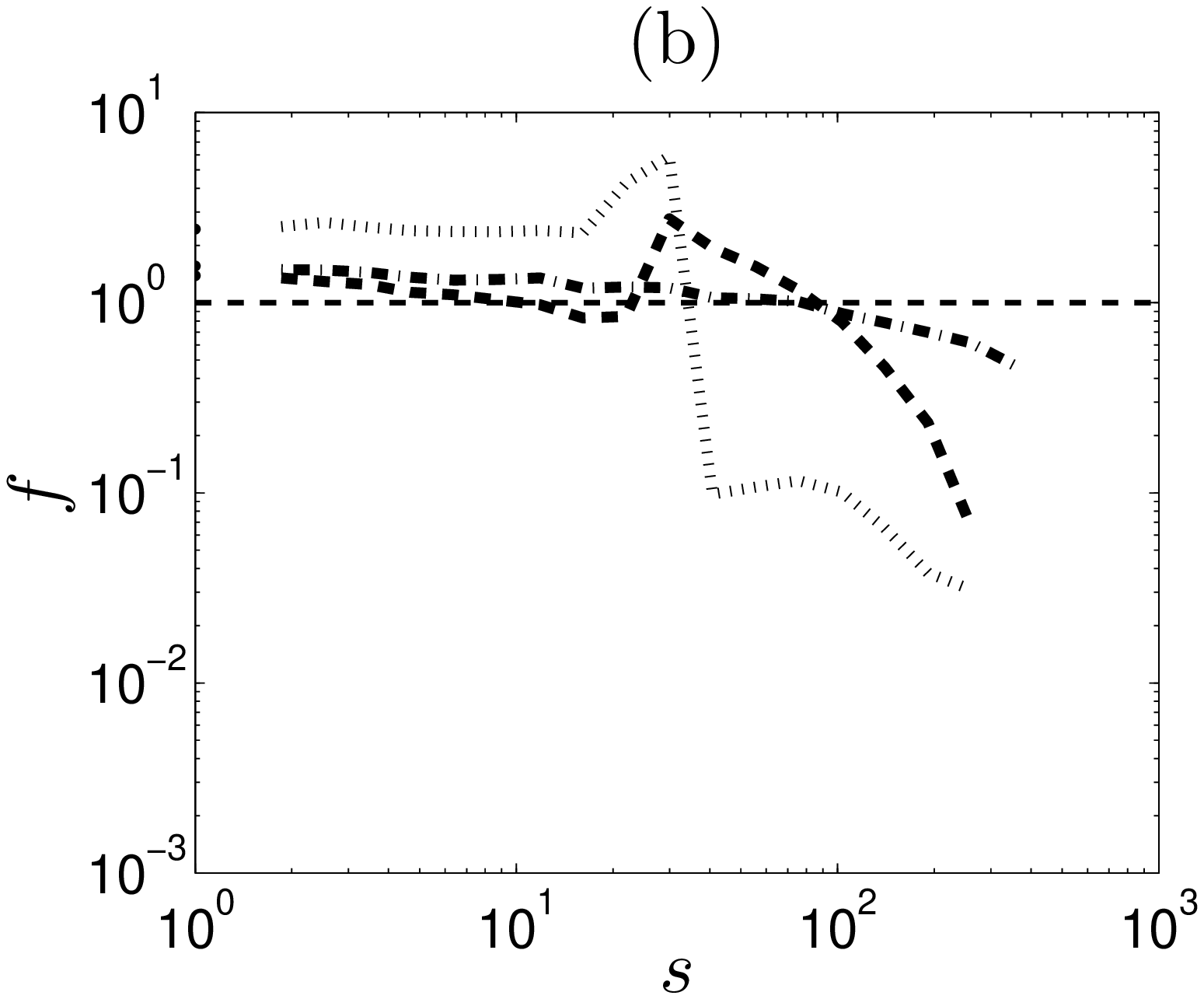} \\
\end{tabular}
\caption{Ratio $f$ between the total number of avalanches in the
controlled and uncontrolled situations for square lattices with
size $L=32$ and for several control schemes. Both in panel (a) as
in (b), the following notation is used to identify parameter
values and adopted control scheme: $(p_I; \mathrm{scheme})$. In
panel (a), $T_I=8$: dashes (0.92; {\it max}), dots (1; {\it min}),
dot-dashes (1; {\it ran}). In panel (b), $T_I=32$: dashes (0.33;
{\it max}), dots (0.78; {\it min}), dot-dashes (0.50; {\it
ran}).}\label{figuraNova}
\end{figure}

Figure \ref{figuraNova} reinforces the results of Figure
\ref{figura3} showing simulations of the DADR's model controlled
by the heuristics {\it max}, {\it min} and {\it ran} for a system
with size $L=32$. While in Figure \ref{figuraNova}(a) $T_I=8$
(small value), in figure Figure \ref{figuraNova}(b) $T_I=32$
(large value). One should note that qualitatively the results are
the same. Furthermore, based on the results of Figure
\ref{figura3}, we are able to say that while in the first case
($T_I=8$) the heuristic {\it max} is closest to the optimality, in
the second case ($T_I=32$) the heuristic ({\it min}) is the one
that it is closest.


\section{Final remarks} \label{sec:conc}

We have introduced a DP approach to control SOC in the DADR model.
Although this framework cannot be applied to large system, it is
quite useful to characterize the optimal solution and evaluate
optimality of other heuristics. The reduction in the number of
large avalanches shown in Fig. 2 is similar to those obtained
in~\cite{cajand10}, where a fixed heuristics was considered. In
that work, no cost was associated with intervention, so that it is
not possible to directly compare results predicted in Fig. 1a to
larger systems. However, the sudden vanishing of $\langle
s\rangle$ at $(C_I/C_A)_T$ suggests that, for heuristic based
control, a similar behavior would be observed if cost is
introduced in the model. Finally, this approach can be the basis
to study approximate sub-optimal approaches in the line
of~\cite{bertsi96}, using for instance reinforcement learning
techniques.

\section{Acknowledgment} The authors are grateful to the Brazilian
agency CNPQ and the National Institute of Science and Technology
for Complex Systems (Brazil) for financial support.

\end{document}